\documentclass[aps,prl,reprint,superscriptaddress,showpacs]{revtex4-1}

\usepackage{graphicx}
\usepackage{dcolumn}
\usepackage{bm}
\usepackage{amssymb}

\begin{document}

\title{Suppression and emergence of granular segregation under cyclic shear}

\author{Matt~Harrington}
\email{mjharrin@umd.edu}
\affiliation{Department of Physics and IREAP, University of Maryland, College Park, Maryland 20742, USA}

\author{Joost~H.~Weijs}
\affiliation{Physics of Fluids Group and J.M. Burgers Centre for Fluid Dynamics, University of Twente, Post Office Box 217, 7500 AE Enschede, Netherlands}

\author{Wolfgang~Losert}
\email{wlosert@umd.edu}
\affiliation{Department of Physics and IREAP, University of Maryland, College Park, Maryland 20742, USA}

\date{\today}

\begin{abstract}
While convective flows are implicated in many granular segregation processes, the associated particle-scale rearrangements are not well understood.  A three-dimensional bidisperse mixture segregates under steady shear, but the cyclically driven system either remains mixed or segregates slowly. Individual grain motion shows no signs of particle-scale segregation dynamics that precede bulk segregation. Instead, we find that the transition from non-segregating to segregating flow is accompanied by significantly less reversible particle trajectories, and the emergence of a convective flow field.  
\end{abstract}     


\maketitle

When mixtures of granular materials are continuously disturbed by external forcing, such as vibration, gravity, rotation, or shear, the grains often separate based on their species properties, such as size or density \citep{SavageLunJFM1988,ClementPRL1993,KnightNagel1993,Khakhar1997,GrayHutter1997,JainLueptowPRE2005,CharlesMorrisGRANMATT2006,GolickDanielsPRE2009,HillFanPRE2010,
MayDanielsPRE2010}.  This phenomenon, known as \emph{segregation}, has been a subject of scientific interest for several decades because of its widespread applicability in both nature and industry, from stratification of avalanche deposits to the nonuniform settling of mixed nuts and cereals.  

\indent
Previous studies have looked at the onset and patterns of segregation, and several theoretical models have been proposed \citep{SavageLunJFM1988,Khakhar1997,GarzoDufty2002,GrayThornton2005}.  A universal feature of the models is gradient-driven segregation flux.  In the specific case of dense shear-driven segregation, however, the relative contributions of gradients in friction, shear, and average kinetic energy (or kinetic temperature) are not well established.  In addition, most candidate models for shear-driven segregation are linear \citep{GrayThornton2005,FanHillNJoP2011}.  A key nonlinearity missing from the current models is granular convection, which can play a vital role in the segregation process \citep{KnightNagel1993,KhosropourBehringerPRE1997,HillFanPRE2010}.  Indeed, the behavior of experimental shear-driven systems deviates from what current models predict \citep{MayDanielsPRE2010}.  

\indent  
In particular, convective flows can lead to the rise of large particles in vibrated beds \citep{KnightNagel1993}.  Even under quasistatic vibrations with very low kinetic temperature and convection, critical size ratios for the rising of large intruders have been observed \citep{ClementPRL1993,LudingJPF1995}.  While this has inspired further work on vibroexcited systems, there is little discussion in the current literature about shear-driven segregation in the absence of kinetic temperature.  

\indent
One distinguishing feature of shear-driven flows is that, unlike vibroexcited systems, they can exhibit flow driven by bulk shear forces, with particle-particle interaction potential energies that are orders of magnitude larger than kinetic energies \citep{KondicBehringerEPSL2004}.  Driven slowly, granular shear flows can then exhibit close to reversible particle trajectories under oscillatory driving \citep{PineGollubNature2005,SlotterbackPRE2012}.  Continuum models of segregation are based on gradients, therefore invariant under reversal of shear direction, and predict the same segregation flux for both steady and oscillatory shears.  Indeed, the absolute size of individual particles relative to the amount of shear does not enter into these models, nor does a characteristic length scale over which gradient-induced segregation manifests.

\indent
In this Letter, we provide insight into these challenges by presenting experimental results from a dense three-dimensional (3D) shear-driven bidisperse granular material.  Using the refractive index matched scanning (RIMS) imaging technique \citep{RIMS2012}, we are able to experimentally determine the trajectories of almost all particles in the mixture as the system is slowly sheared, both steadily and cyclically, in a split-bottom geometry.  We directly measure the bulk segregation patterns under steady and cyclic shear and characterize segregation at the micro- and mesoscale by quantifying the reversibility of particle rearrangements and examining the resulting secondary flows of the system.

\begin{figure}
\includegraphics[width=\linewidth]{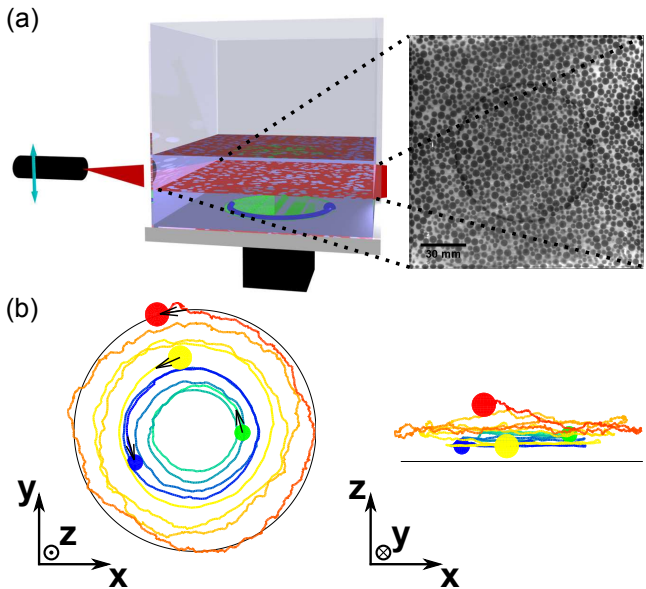}
\caption{\label{fig:ExpSetup} (a) The split-bottom shear cell and a sample horizontal cross section (an aligned second laser sheet and a high sensitive camera that captures images from the top down are not pictured). (b) Sample trajectories of a small (light green to dark blue) and large (light yellow to dark red) grain under steady shear.  The black circle represents the shearing disk.}
\end{figure}

\emph{Setup and procedure.---} The granular system is a dense bidisperse mixture of polymethyl methacrylate spheres, with small and large diameters of $D_S = 3.175$ mm and $D_L = 4.7625$ mm, respectively.  The mass ratio between the two species of grains ($\frac{M_S}{M_L} = \frac{16}{27}$) is selected such that the number ratio of small to large grains $\frac{N_S}{N_L} = 2$.  The sides of the square tank containing the pile are 15 cm long, and the pile height is 4.6 cm, approximately equal to the radius of the shearing disk $R_s = 4.5$ cm, as well as $10 D_L$.   

\indent
The grains are immersed in an index matched interstitial fluid, Triton X-100.  The index-matching allows light from two laser sheets, mounted on either side of the tank, to pass through undeflected.  We also add a laser dye, Nile Blue 690 Perchlorate, that fluoresces at the laser wavelength (635 nm), as well as a small amount of hydrochloric acid to stabilize the mixture.  With the laser sheets aligned at a particular height, a high-sensitivity Sensicam camera captures a horizontal cross section of the pile from the top down.  The laser sheets and camera move incrementally along stepper motors, with cross sections taken about every 200 $\mu$m in height.  All of the cross sections together construct a full 3D image of the pile.  An illustration of our setup is shown in Fig.~\ref{fig:ExpSetup}(a).  

\indent
We consider the response of the granular system to both steady and cyclic shear, with amplitudes of 10$^{\circ}$ and 40$^{\circ}$.  The shearing disk is separated from the rest of the tank floor and rotates at a rate of 1 mrad/s, with static 3D images captured every 2$^{\circ}$.  At this shearing rate, the flow profiles of the pile resemble that of a dry pile and are also rate independent \citep{DijksmanPRE2010}.  In addition, the circular split-bottom geometry allows for the formation of wide and robust shear zones far from the system walls \citep{FenisteinVanHeckeNature2003,FenisteinVanHeckePRL2004}.

\indent
When an experiment is complete, particle center positions are determined using a 3D-adapted convolution kernel \citep{TsaiGollubPRE2004}, which also distinguishes between large and small grains.  In each frame, we are capable of extracting the position of at least 95\% of particles, with a particle center resolution of about 100 $\mu$m.  Then, a Lagrangian predictive particle tracking algorithm \citep{OuelletteExpFluids2006} identifies individual particles and their trajectories.  Examples of individual particle trajectories, for a large and small grain under steady shear, are shown in Fig.~\ref{fig:ExpSetup}(b).

\begin{figure}
\includegraphics[width=\linewidth]{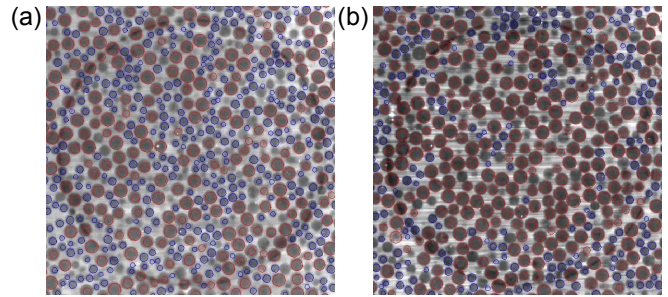}
\caption{\label{fig:cross-sections} Cross sections of the granular pile at 3.8 cm above the shearing disk, or about 2 large grain diameters from the free surface, (a) in the initial mixed state and (b) after 29.3 revolutions of the shearing disk.  Large red and small blue circles represent large and small grains, respectively.}
\end{figure}

\begin{figure}
\includegraphics[width=\linewidth]{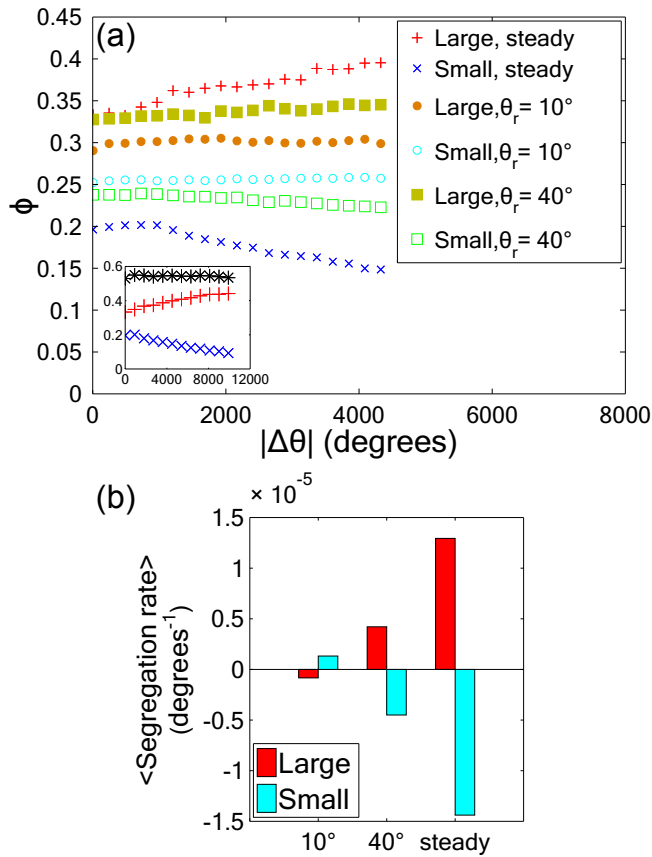}
\caption{\label{fig:vfraction} (a) Volume fractions of large and small grains under all shearing procedures considered: steady (red +'s and blue $\times$'s), $\theta_{r} = 10^{\circ}$ (filled orange and empty cyan circles), and $\theta_{r} = 40^{\circ}$ (filled gold and empty green squares) in the top 60\% of the pile, directly above the shearing disk.  Inset: Same as (a), but showing only steady shear and including total volume fraction (black asterisks).  (b) The average rate of change of $\phi$ of large and small grains for $|\Delta\theta| \geq 1000^{\circ}$.}
\end{figure}

\emph{Bulk segregation.---} A standard segregation pattern that is seen in many polydisperse mixtures is the Brazil nut effect (BNE), in which large particles rise to the top of the pile, leaving smaller sizes to settle toward the bottom.  In this slowly sheared system, a pattern similar to the BNE is observed under steady shear.  Fig.~\ref{fig:cross-sections} shows how a cross section at a height of 3.8 cm (eight large grain diameters) becomes primarily populated with large particles after almost 30 revolutions of steady shear.  This height is close to the free surface of the pile, only two large grain diameters away.  This effect is also indicated by the sample trajectories shown in Fig.~\ref{fig:ExpSetup}(b).  The pile as a whole rearranges such that large particles inhabit a cylindrical region at the top 60\% ($\approx V_L/V_{\mathrm{total}}$, where $V_L$ is the total volume of large grains and $V_{\mathrm{total}}$ is the total volume of \emph{all} grains) of the pile, directly above the shearing disk.  To quantify bulk segregation, we calculate the volume fraction of large and small grains in this region of interest.  The inset of Fig.~\ref{fig:vfraction}(a) illustrates that over about 30 rotations of the shearing disk, the pile is continually segregating, with a total volume fraction that remains consistent throughout.  

\indent
Under oscillatory shear, however, segregation is not necessarily observed.  The results for bulk segregation under steady shear and the two oscillatory amplitudes considered are summarized in Fig.~\ref{fig:vfraction}.  When the amplitude is 10$^{\circ}$, the volume fractions of large and small particles appear to remain constant throughout.  This indicates that the system is not segregating over the time interval measured.  Individual particle trajectories are consistent with this observation, as discussed in the following section.

\indent
For sufficiently large amplitudes of oscillatory shear, the pile does slowly segregate.  For an amplitude of 40$^{\circ}$, the rate of segregation is smaller than under steady shear.  However, segregation is apparent over the measured time period and the associated changes in volume fraction are in accord with BNE.   

\emph{Microscale reversibility.---} The RIMS technique allows us to also probe segregation microscopically by considering particle contact networks and mean square displacements (MSDs) within the shear zone of the split-bottom geometry, where most rearrangements occur \citep{FenisteinVanHeckeNature2003,FenisteinVanHeckePRL2004,SlotterbackPRE2012}.

\indent
In order to verify that the granular material is segregating, we must see that grain motion is irreversible in terms of both grain-to-grain contacts and displacements.  The pile starts in a highly mixed state, so each individual grain touches a combination of small and large grains.  Under segregation, particle contacts must be constantly changing, in order to form new contacts with similar particles.  At the same time, segregation requires significant and irreversible motion within the pile, so that an increasingly segregated state develops.   

\indent
We determine grain contacts by finding particle pairs that, within some distance cutoff, are increasingly likely to move perpendicular to each other by sliding or rolling, as described by Herrera \textit{et al}~\citep{HerreraPRE2011}.  Three contact length cutoffs are defined: large-large, large-small, and small-small.  MSDs are simply calculated from the average square displacement of grains from an initial reference state to a new state at a later time.

\begin{figure}
\includegraphics[width=\linewidth]{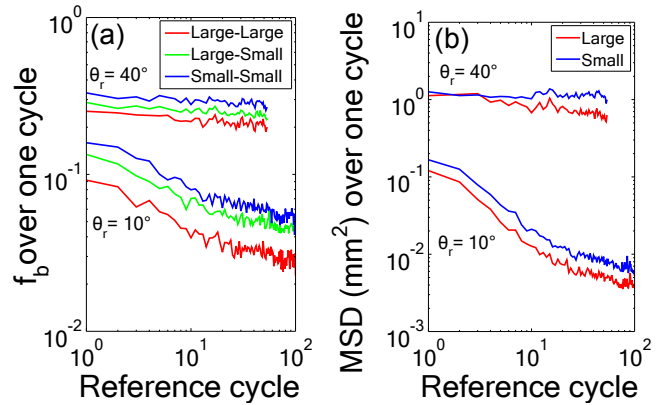}
\caption{\label{fig:fb_msd_onecyc} (a) Fraction of broken grain contacts and (b) MSDs over a single cycle versus that particular cycle number.  
}
\end{figure}

\indent  
Fig.~\ref{fig:fb_msd_onecyc}(a) shows the fraction of broken links at the end of a single shear cycle, with respect to the reference network taken at the end of the previous cycle.  For all three contact types, there is a decreasing fraction of contacts that are broken over time for the 10 degree experiment; however, the fraction of broken links for the 40 degree experiment remains fairly steady from cycle to cycle.  Similar trends are observed if we measure the MSDs of large and small grains over individual cycles, as shown in Fig.~\ref{fig:fb_msd_onecyc}(b).  This distinguishing behavior of reversible and irreversible flows was also seen by Slotterback \textit{et al} in a monodisperse system with particle size $D_L$, with the same shear amplitudes \citep{SlotterbackPRE2012}.  This indicates that in this split-bottom geometry, there is a connection between a shear-driven granular material undergoing irreversible flows and having the capacity to segregate.

\begin{figure*}
\includegraphics[width=510pt]{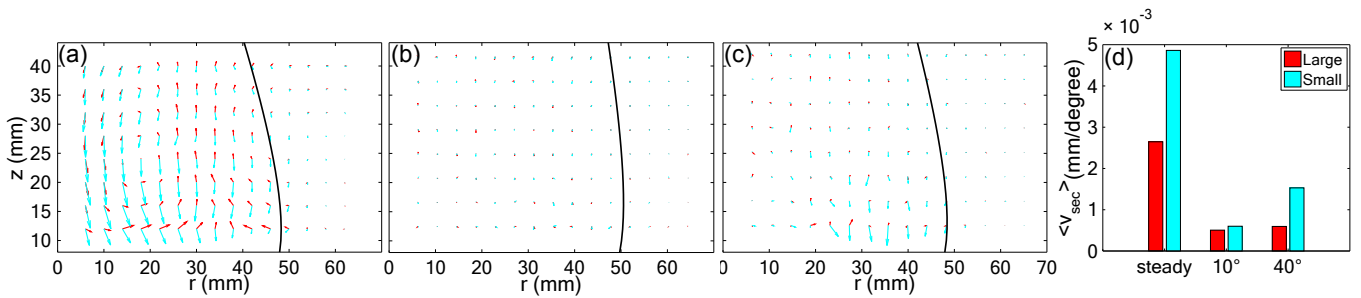}
\caption{\label{fig:largesmallflows_all} Time- and $\theta$-averaged flows of large (dark red arrows) and small (light cyan arrows) grain flows plotted together for all shearing procedures considered: (a) steady, (b) $\theta_r = 10^{\circ}$, and (c) $\theta_r = 40^{\circ}$.  Vectors are scaled such that the grid spacing corresponds to $5.6 \times 10^{-3}$ mm/degree shear. (d) Average magnitude of secondary flow vectors within the shear zone outer boundary [black curves in (a)--(c)].}
\end{figure*}

We have described the bulk segregation patterns we see under steady and cyclic shear and linked our segregating and nonsegregating regimes to changes in particle-scale reversibility.  However, segregation requires rearrangements on much larger than particle scales.  To capture such larger-scale rearrangements, we are motivated by the prevalence of convective flows in other segregating systems to examine secondary flow profiles.

\begin{figure}
\includegraphics[width=\linewidth]{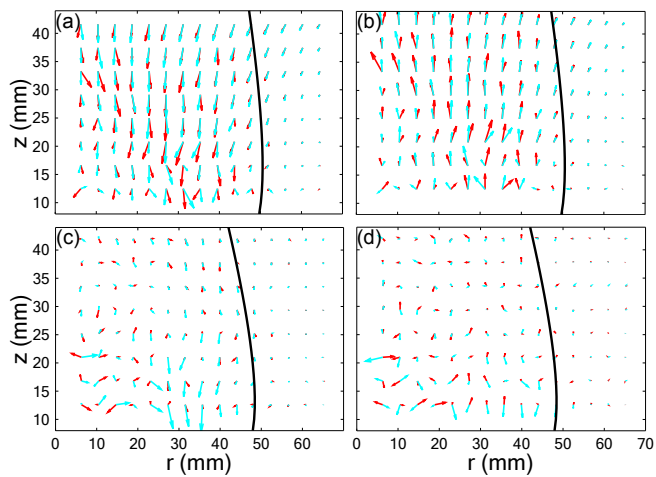}
\caption{\label{fig:forback} For $\theta_r = 10^{\circ}$, reversible secondary flows can be seen when averaging over (a) positive and (b) negative strain.  For $\theta_r = 40^{\circ}$, such a strong distinction cannot be made between (c) positive and (d) negative strain.  Flow vectors are shown on the same scale as Fig.~\ref{fig:largesmallflows_all}.}
\end{figure}

\indent
\emph{Secondary flows.---} The motion of the grains is primarily in the $\theta$ direction, as grains near the bottom are driven locally by the rotation of the shearing disk.  However, the grains are also free in move in the other two cylindrical coordinates $r$ and $z$.  The time- and azimuthal-averaged motions of grains in $r$ and $z$ are referred to as \emph{secondary flows}.  Simulations in the split-bottom geometry suggest that secondary flows are a key component of the segregation process \citep{HillFanPRE2010}. 

\indent
For steady shear, the secondary flows for large and small particles are shown in Fig.~\ref{fig:largesmallflows_all}(a).  The flow profiles of the two species are vastly different; while the large particles form two distinct convection rolls, the small particles primarily drift toward the bottom.  There is a net downward flux of small particles above the shearing disk, which is in line with the bulk segregation pattern.  Also shown in Figs.~\ref{fig:largesmallflows_all}(a)-\ref{fig:largesmallflows_all}(c) is the outer border of the shear zone, which is determined from the shape of the azimuthal, or primary, flow profiles \citep{FenisteinVanHeckeNature2003,FenisteinVanHeckePRL2004}.  The secondary flows exist within this boundary, suggesting that the segregating region is necessarily determined by the size of the particles and the shearing disk.  Furthermore, the symmetry of our system indicates that the magnitude and direction of the flow fields are independent on the direction of steady shear.

\indent
Under oscillatory shear, the secondary flow profiles are dependent on the shear amplitude.  For an amplitude of 10$^{\circ}$, the secondary flows of both species are suppressed, as illustrated in Fig.~\ref{fig:largesmallflows_all}(b).  This is to be expected, given the increasing reversibility of the system over time, and is also reflected when averaging over only positive and negative strain.  The motion illustrated in Figs.~\ref{fig:forback}(a) and \ref{fig:forback}(b) indicates that the system merely exhibits alternating phases of compaction and dilation.  When the shear amplitude is increased to 40$^{\circ}$, such reversibility in the fabric is not observed due to overall trajectory irreversibility.  While granular convection remains mostly undeveloped for large particles, small particles, particularly those close to the shearing disk, drift downward, as shown in Figs.~\ref{fig:largesmallflows_all}(c),~\ref{fig:forback}(c), and~\ref{fig:forback}(d).  For higher amplitudes, as irreversibility increases, we expect to see stronger downward drifts of small particles, as well as granular convection of large grains to approach that of steady shear.

\emph{Discussion.---} Using the RIMS technique, we image and track individual grains in a dense 3D bidisperse granular mixture under the influence of shear in a split-bottom geometry.  We observe that in a regime of slow, rate-independent steady shear, the system segregates with large particles occupying the top portion of the pile directly above the shearing disk.  When the system is sheared cyclically, there is a critical shear amplitude below which segregation is suppressed.  

\indent
We find that bulk segregation may be linked to the microscopic reversibility of small and large grain trajectories.  Under 10$^{\circ}$ oscillatory shear, trajectories of small and large grains are increasingly reversible over time, as observed structurally from their grain contacts and spatially from their MSDs.  However, at 40$^{\circ}$ oscillatory shear, the degree to which the large and small grain trajectories are reversible remains steady over all cycles.  These trends of microscopic reversibility are very similar to observations of the monodisperse system \citep{SlotterbackPRE2012}, indicating that they are not due to any additional forces, unique to granular mixtures, that drive segregation.

\indent
Finally, we investigate the secondary flow profiles brought about by the (ir)reversibility of the system.  Under steady shear, we find clear convective flows of large grains, which help drive the segregation mechanism by allowing large grains to settle at the top of the pile, while a net flux of small grains falls toward the bottom of the container.  These flows are also confined within the outer boundary of the shear zone, suggesting that the size and shape of these flows are dependent on the system geometry.  For small amplitude oscillatory shear, however, secondary flows are reversible between positive and negative strain, indicating that the system fabric retains its original configuration.  Above some critical shear amplitude, the fabric breaks sufficiently and the flow profiles are less distinguishable between positive and negative strain.  While convection of large grains has not completely formed, there is a clear downward drift of small grains near the shearing disk.

\indent
In this system, we observe distinct rates of segregation for steady and adequately large amplitude oscillatory shear.  However, we cannot immediately attribute these rates to relevant system parameters, such as shear amplitude, diffusion coefficients, or gradients of local strain.  A recent study also indicates the presence of gravity is a determinant of granular convection \citep{MurdochPRL2013}, which would be relevant for segregation rates in astrophysical applications.  Future work on this system should focus on probing the functional dependence of segregation rate on these variables, which is key to determining a general predictive model for shear-driven segregation \citep{Thornton2011}.  

\indent
We are encouraged by the apparent existence of a critical shear amplitude that brings about segregation.  In fact, a critical strain of similar magnitude also exists for fracture \citep{HerreraPRE2011} and irreversibility \citep{SlotterbackPRE2012}, suggesting that this granular system may have a universal length scale that drives inherently bulk rearrangements, yet arises directly from particle-scale dynamics.  This length scale could be instrumental in the development of not only segregation models but also predictive constitutive models for granular flows that rely on a characteristic length for cooperative rearrangements \citep{KamrinPRL2012}.

We would like to acknowledge enlightening discussions with Mitch Mailman, Kerstin Nordstrom, and Anthony Thornton, as well as technical and analytic assistance from Steven Slotterback.  Financial support for this work came from the National Science Foundation (DMR0907146) and the Defense Threat Reduction Agency (HDTRA1-10-0021).

\bibliography{Harrington_Bidisperse_Bib}

\end{document}